\documentclass[twocolumn]{aastex}
\usepackage{amsmath}
\usepackage[figuresright]{rotating}

\begin{document}
%
%
%
%
\title{Filament Mass Losses Forced by Magnetic Reconnection in the Solar Corona}
%
%
%
%
\author[0000-0003-4023-9887]{Craig D. Johnston}
\affiliation{Department of Physics and Astronomy,
George Mason University,
Fairfax, VA 22030, USA}
\affiliation{Heliophysics Science Division,
NASA Goddard Space Flight Center,
Greenbelt, MD 20771, USA}
\correspondingauthor{Craig D. Johnston}
\email{craig.d.johnston@nasa.gov, cjohn44@gmu.edu}

\author[0000-0002-1198-5138]{Lars K. S. Daldorff}
\affiliation{Department of Physics,
Catholic University of America,
Washington, DC 20064, USA}
\affiliation{Heliophysics Science Division,
NASA Goddard Space Flight Center,
Greenbelt, MD 20771, USA}

\author[0000-0003-1522-4632]{Peter W. Schuck}
\affiliation{Heliophysics Science Division,
NASA Goddard Space Flight Center,
Greenbelt, MD 20771, USA}

\author[0000-0002-4459-7510]{Mark G. Linton}
\affiliation{Space Science Division,
Naval Research Laboratory,
Washington, DC 20375, USA}

\author[0000-0001-9642-6089]{Will T. Barnes}
\affiliation{Department of Physics,
American University,
Washington, DC 20016, USA}
\affiliation{Heliophysics Science Division,
NASA Goddard Space Flight Center,
Greenbelt, MD 20771, USA}

\author[0000-0003-0072-4634]{James E. Leake}
\affiliation{Heliophysics Science Division,
NASA Goddard Space Flight Center,
Greenbelt, MD 20771, USA}

\author[0000-0002-0461-3029]{Simon Daley-Yates}
\affiliation{School of Physics and Astronomy, 
University of St Andrews,  
St Andrews,  KY16 9SS, UK}

%
%
%
%
\begin{abstract}

Recent observations of the solar atmosphere in cool extreme ultraviolet (EUV) lines have reported the prevalence of coronal rain falling from coronal cloud filaments that are associated with the magnetic dips of coronal X-point structures. These filaments mysteriously appear as clouds of mass in the corona that subsequently shrink and disappear due to mass losses that drain as coronal rain along arced field lines. Using a two and a half dimensional, magnetohydrodynamic model, we investigated evaporation–condensation as the formation mechanism of the subset of coronal cloud filaments that form above coronal X-points. Our simulation included the effects of field-aligned thermal conduction and optically thin radiation and used the state-of-the-art Transition Region Adaptive Conduction (TRAC) method to model the formation, maintenance, and mass loss of a filament above a coronal X-point.  This paper presents a physical model that demonstrates magnetic reconnection as a filament loss mechanism, producing hybrid filament/coronal rain via mass losses through the X-point. A detailed analysis of how the mass of the filament forces the field to reconnect is also presented, revealing three phases that characterize the evolution of the reconnecting current sheet and associated mass losses. We conclude that the formation of certain coronal cloud filaments and subsequent mass losses via coronal rain can be explained by the evaporation–condensation model combined with filament mass losses forced by magnetic reconnection. We also report that rebound shocks generated by the impact of coronal rain condensations on the chromosphere together with retractive upflows can cause upward propagating condensations to form through a dynamic thermal runaway process. 

\end{abstract}
%
%
%
%
%
%
\keywords{Solar corona (1483); Solar magnetic fields (1503); Solar filaments (1495); Solar prominences (1519); Solar magnetic reconnection (1504);  Magnetohydrodynamics (1964)}
%
%
%
%
\section{Introduction} \label{sec:intro}

Solar filaments are the signature of cool ($\approx 10^4$~K), dense ($\approx 10^{17}$ m$^{-3}$) plasma that is suspended in the much hotter, more rarefied solar corona \citep[for a comprehensive review, see e.g.,][]{Labrosse2010,Mackay2010,Parenti2014,Gibson2018}. On the bright solar disk, filaments appear as dark structures with the plasma in absorption. Above the limb, these structures appear in emission as bright \lq\lq prominences\rq\rq\ in cool extreme ultraviolet (EUV) lines.

Filaments are always found in the corona above magnetically sheared polarity inversion lines (PILs), where \lq\lq magnetic shear\rq\rq\ refers to the component of the magnetic field along the PIL that is built up by shearing and/or converging magnetic footpoint motions \citep{Martin1998}. Three main classes of filament have been identified in observations: quiescent, intermediate (combined) and active region filaments \citep[][]{Mackay2010,Mackay2015}. Quiescent filaments form in quiet regions of the Sun and around the polar crown, intermediate filaments form around the borders of active regions, and active region filaments form within the centers of activity nests of multiple pairs of sunspots. Typically, quiescent and intermediate filaments are larger, more stable structures with longer lifetimes (weeks to months) compared to active region filaments, which are generally unstable with a short lifetime (hours to days).

Associated with each of these filament classes are two magnetically different types of filament: channel filaments and coronal cloud filaments \citep[][]{Martin2015,Martin2016}. Channel filaments owe their existence to the presence of an external magnetic environment known as a \lq\lq channel\rq\rq\  whose central field is parallel with the filament spine \citep{Gaizauskas1997,Gaizauskas2001}, which runs horizontally along the top of the filament. These filaments are well connected to the chromosphere along their spines or at barbs that protrude from the side of the filament, and are generally agreed to be precursors to major eruptive events. In contrast, coronal cloud filaments form at isolated heights in the corona \citep[][]{Leroy1972,Allen1998,Liu2012,Liu2014,Liu2016} and are rarely seen to erupt. Instead, they mysteriously appear as clouds of mass in the corona that shrink and disappear within a few hours to a day, draining along well-defined arcs as coronal rain.

This type of rain falling from coronal cloud filaments is referred to as \lq\lq hybrid filament/coronal rain\rq\rq \citep[][]{Liu2012,Liu2016,Antolin2022} to distinguish it from the quiescent coronal rain \citep{Antolin2012,Auchere2018,Froment2020} and flare-driven rain \citep{Foukal1978,Jing2016,Brooks2024} that are also observed in active regions. While all three share many similar characteristics such as a small, clumpy morphology and a cool, dense chromospheric core, these distinctions are important because each type of rain has a unique formation process \citep{Antolin2022}. In particular, hybrid filament/coronal rain is directly linked to the magnetic topology and stable filaments, and thus magnetic dips appear necessary for its existence. This is not the case for either quiescent or flare-driven rain. Moreover, coronal rain has also been observed on other stars \citep[e.g.,][]{Namekata2021,Namekata2022}, and recent work by \cite{Yates2023} demonstrated that magnetic dips are not necessary for rain on rapidly rotating young stars. 

Despite over 150 years of observations \citep{Secchi1875, Secchi1877}, the detailed formation and subsequent draining of coronal cloud filaments is still not fully understood. The main questions that need to be addressed are the source of the mass, the type of magnetic structure that supports the mass, the role of reconnection, and the regulation of the mass losses that drain as hybrid filament/coronal rain. 

\cite{Martin2016} suggested that the mass for many coronal cloud filaments comes from previously erupted filaments. In this scenario, part of the erupted mass falls back to the Sun where it is collected in magnetic dips, before slowly draining down from the clouds as coronal rain.
However, this explanation strains credulity because the formation of cloud filaments should then depend strongly on the rate of nearby eruptions.

Indeed, recent observations using the Atmospheric Imaging Assembly \citep[AIA;][]{Lemen2012} on board the Solar Dynamics Observatory \citep[SDO;][]{Pesnell2012} point to other possible sources of mass, in particular, they have reported the prevalence of coronal rain falling from cloud filaments that are associated with the magnetic dips of coronal X-point structures \citep[][]{Li2018,Li2019,Chen2022}. This draining of hybrid filament/coronal rain has been observed in active regions and the quiet Sun. Furthermore, the fact that such cloud formation and rain events are seen to occur repeatedly over several days \citep{Li2019} significantly supports evaporation–condensation as the mass accumulation mechanism, invoking the thermal non-equilibrium (TNE) process through sustained heating at low coronal altitudes \citep{Antiochos1991,Antiochos1999,Antiochos2000,Karpen2001,Karpen2003,Karpen2005,
Xia2011,Xia2012,Luna2012,Mikic2013,Xia2016,Froment2018,Johnston2019b} to form the clouds. 
In particular, the evaporation–condensation model provides a simple explanation for the replenishment of the filament mass losses so that the filament can persist for several days.
This scenario is further supported by the recent discovery that TNE cycles can also occur on magnetically open field lines \citep{Scott2024}.

Numerical modeling of filament formation in the solar corona has a long history, but remains computationally challenging because of the requirement to account for field-aligned thermal conduction, optically thin radiation and chromospheric evaporation. 
Incorporating these additional physics into a multi-dimensional magnetohydrodynamic (MHD) model of a filament requires resolving (1) the transition region (TR) separating the cool chromosphere ($\approx 10^4$~K) from the hot corona ($\approx 10^6$~K), and (2) any dynamically forming filament-corona transition regions (FCTRs) associated with the filament plasma, which separate the cool, coronal condensations ($\approx 10^4$~K) from the surrounding hot atmosphere ($\approx 10^6$~K).

There are two main consequences of poorly resolving the TR and FCTRs. The first is that under-resolving the TR leads to coronal densities that are artificially low \citep{Bradshaw2013,Johnston2019a}. 
This happens because instead of passing through the TR in a series of small steps, the downward heat flux is forced to jump across an under-resolved TR in a large step.
The large step then results in an over-estimation of the radiation since the chromosphere is denser than the under-resolved region above, leaving little energy left to drive evaporation \citep{Bradshaw2013,Johnston2017a,Johnston2017b}. \cite{Johnston2019b} demonstrated that this under estimation of the evaporation can suppress the occurrence of condensations in the corona and subsequent TNE cycles \citep{Kuin1982,Froment2018,Winebarger2018,Klimchuk2019}. The second consequence is that under resolving the FCTRs leads to discrepancies in the maintenance of condensations and their collective filament characteristics, e.g., temperatures, densities and lifetimes \citep{Johnston2019b}.

Therefore, dense uniform meshes or adaptive re-gridding have historically been required to resolve the TR and FCTRs in numerical simulations of filament formation. This has necessitated the use of two mutually exclusive oversimplified approaches. 
The first approach models the multi-dimensional evolution of sheared magnetic fields in the solar atmosphere with MHD codes while neglecting the details of the thermodynamics and mass exchange necessary for the dynamic formation of condensations. Consequently, these MHD models assume that \lq\lq magnetic shear\rq\rq\ implies a filament \citep{Antiochos1994,Knizhnik2017a,Knizhnik2017b,Gibson2018} and/or that magnetic dips above the PIL imply a filament \citep{DeVore2000,Aulanier2002,Lionello2002,DeVore2005,Gibson2006}.
The second approach employs a hydrodynamic model of the detailed field-aligned thermodynamics and mass exchange to capture the formation of condensations, but neglects coupled thermodynamic processes across field lines and any evolution of the underlying magnetic structure. The state-of-the-art for this approach involves extracting \lq \lq one-dimensional\rq\rq\ field lines from an MHD model of the solar atmosphere \citep{Luna2012,Guo2022}. These field lines are then used to geometrically parameterize the magnetic field in hydrodynamic and radiation calculations with dense spatial grids (e.g., 100~m)  to resolve the TR and any dynamic FCTRs that form \citep{Bradshaw2013,Johnston2019b,Mason2023}. Both of these approaches are unrealistic approximations because the formation of cool, dense filaments in the hot corona is expected to involve the interplay between magnetic and thermodynamic evolution \citep{Keppens2014,Johnston2021}.

Recently, \cite{Johnston2019a} and \cite{Johnston2020} demonstrated that these challenges are overcome, without high spatial resolution, by using the Transition Region Adaptive Conduction (TRAC) method. This is achieved by enforcing certain conditions on the parallel thermal conductivity $\kappa_\parallel(T)$, radiative loss $\Lambda(T)$, and heating $Q(T)$ rates that are due, in their original form, to \cite{Lionello2009} and \cite{Mikic2013}, and were subsequently extended by \cite{Johnston2020, Johnston2021} for use in one-dimensional hydrodynamic and multi-dimensional MHD models. These conditions act to broaden any unresolved parts of the TRs. \cite{Johnston2020} then also showed that the TRAC modifications allow the TRs to be modeled in simulations with computationally manageable grid sizes of order 100~km because the TRAC method (1) preserves the energy balance in the TR and any dynamically forming FCTRs, and (2) conserves the total amount of energy that is delivered to the chromosphere or condensations, consistent with fully resolved models. The TRAC-enabled simulations gave peak density errors of less than $5\%$ and captured the period associated with the formation and maintenance of TNE condensations to within $10\%$, whereas without TRAC, in the equivalent coarse resolution simulations, the errors can be up to 75\% \citep[see, e.g.,][]{Johnston2017a,Johnston2020,Johnston2021} and TNE is suppressed \citep{Johnston2019b}. 

In the work presented in this paper, we take full advantage of the MHD TRAC method \citep{Johnston2021}, in order to (1) simulate how cool, dense plasma collects in a filament above a coronal X-point and (2) investigate how magnetic reconnection then regulates the coronal rain. 
We conclude that the formation of certain cloud filaments and subsequent coronal rain can be explained by the evaporation–condensation model combined with filament mass losses forced by magnetic reconnection. 
In Section 2, we describe the key features of our numerical model. Section 3 discusses our results, focusing on the formation of the filament, the draining of hybrid filament/coronal rain and the role of reconnection. Finally, in Section 4, we summarize our results and discuss comparisons with previous simulations and observations, before drawing our conclusions.

%
%
%
%
\section{Numerical Model} \label{sec:model}

\subsection{Governing Equations}

To model the formation and maintenance of a filament, including the interplay between its magnetic and thermodynamic evolution, the following set of MHD equations, which incorporate gravitational stratification, field-aligned thermal conduction and optically thin radiation, are solved numerically using version 3.3 of the Lagrangian Remap (LaRe) code \citep{Arber2001}:
  \begin{eqnarray}
    \frac{\partial\rho}{\partial t}  
    + \nabla \cdot (\rho {\bf v})
    = 0;
    \hspace{5cm}
    \label{Eqn:mhd_continuity}
    \\[1mm]
    \rho \frac{D{\bf v}}{Dt}
    = - 
    \nabla P + \rho {\bf g} + {\bf j \times B} + 
    {\bf F}_{\textrm{visc}};
    \hspace{2.5cm}
    \label{Eqn:mhd_motion}
    \\[1mm]
    \frac{\partial {\bf B}}{\partial t}
    = 
    \nabla \times ({\bf v \times B});
    \hspace{5cm}
    \label{Eqn:mhd_induct}
    \\[1.5mm]
    \rho \frac{D \epsilon}{Dt}
    = -P\nabla \cdot{\bf v} 
    + Q_{\textrm{visc}} \!
    -\nabla \cdot {\bf q}  
    -  n^2 \Lambda(T)
    \! + Q_{\textrm{cor}};
    \hspace{0.5cm}
    \label{Eqn:mhd_ee}
    \\[1.5mm]
    P = \rm{k_B} n T.
    \hspace{6.5cm}
    \label{Eqn:mhd_gas_law}
  \end{eqnarray} 
In these equations, $\rho$ is the mass density,  
${\bf v}$ is the velocity, 
$P$ is the gas pressure, 
${{\bf g}=-274 \textrm{ ms}^{-2}}{\bf \hat{z}}$ 
is the gravitational acceleration,
${\bf j=(\nabla \times B})/\mu_0$ is the current density and
${\bf B}$ is the magnetic field.
${\bf F}_{\textrm{visc}}$ represents the viscous force that is associated with shock viscosity terms which are included to ensure numerical stability, as described in \cite{Arber2001}, and $Q_{\textrm{visc}}$  is the corresponding viscous heating term.
The specific 
internal energy density is given by $\epsilon=P/(\gamma-1) \rho $
(where $\gamma=5/3$ is the ratio of specific heats), 
$n$ is the number density ($n=\rho/1.2\, \rm{m_p}$, where $\rm{m_p}$ is the
proton mass),
$\rm{k_B}$ is the Boltzmann constant and
$T$ is the temperature.

An explicit resistivity, $\eta$, is not included in the induction Eq.~\eqref{Eqn:mhd_induct}.
Therefore, LaRe’s finite numerical resistivity facilitates the magnetic reconnection that occurs in the current sheets,  which form dynamically during the filament formation process.

Furthermore, the thermodynamic terms that play a key role in filament formation are included in the energy Eq.~\eqref{Eqn:mhd_ee}. The heat flux vector ${\bf q}$ is based on the \cite{Braginskii1965} formulation  in the presence of a magnetic field, which recovers the field-aligned Spitzer-H{\"a}rm parallel thermal conductivity in the strong field limit \cite[]{Spitzer1953}.  
Special treatment of the weak field limit is also incorporated so that the thermal conductivity becomes isotropic at magnetic null points \citep{Johnston2021,Johnson2024}.
The radiative loss function $\Lambda(T)$ of an optically thin plasma is approximated using the 
piecewise continuous function defined in
\cite{Klimchuk2008}.
$Q_{\textrm{cor}}$ is an imposed ad-hoc coronal heating function that includes a small background component ($Q_{\textrm{bg}}$) and a stronger localized footpoint heating term ($Q_{\textrm{fp}}$) so that $Q_{\textrm{cor}}=Q_{\textrm{bg}}+Q_{\textrm{fp}}$.

The TRAC method developed by \cite{Johnston2020,Johnston2021} accurately captures the thermodynamics of the computationally demanding TR without the need for high spatial resolution. In this paper, we used the localized MHD formulation of TRAC presented in \cite{Johnston2021} because of its unique ability to (1) accurately treat the TR and any dynamically forming FCTRs associated with the filament plasma, and (2) automatically account for changes in magnetic field line connectivity. This formulation thus facilitated simultaneous modeling of the filament's dynamic magnetic field and thermodynamic evolution of its plasma.

\subsection{Initial Conditions \& Boundary Conditions}

The simulation is initialized with a two and a half dimensional (2.5D) sheared, quadrapolar magnetic field on a two dimensional (2D) computational domain of extents 
$X \times Z = [-40, 40] \times [0, 80] ~\textrm{Mm}$.
A uniform numerical grid comprised of $512 \times 512$ points is used to resolve this domain, giving a spatial resolution of approximately 150~km in both the horizontal and vertical directions.

Marginally dipped magnetic field lines are found at coronal heights above an X-point that is also located in the corona for the quadrapolar field given by:
  \begin{eqnarray}    
    B_x
    = -B_0\cos(nx) \exp(-mz)\cos(\theta) \hspace{1cm}
    \nonumber
    \\
    + B_1\cos(3nx) \exp(-3mz)\cos(\theta); \hspace{1cm}
    \label{Eqn:Bx}
    \\[1.5mm]
    B_y
    = -B_0\cos(nx) \exp(-mz)\sin(\theta) \hspace{1cm}
    \nonumber
    \\
     + B_1\cos(3nx) \exp(-3mz)\sin(\theta); \hspace{1cm}
    \label{Eqn:By}
    \\[1.5mm] 
    B_z
    = + B_0\sin(nx) \exp(-mz) \hspace{2cm}
    \nonumber
    \\
    - B_1\sin(3nx) \exp(-3mz), \hspace{2cm}
    \label{Eqn:Bz}
  \end{eqnarray} 
where $n=\pi/L_x$, $m=n\cos(\theta)$ and $L_x=80$~Mm is the horizontal size of the domain. We take $B_0=3\times 10^{-3}$~T, $B_1=7.5\times 10^{-3}$~T and set $\theta=60^\circ$ to obtain a coronal quadrapole that is strongly sheared, making a $30^\circ$ shear angle ($0^\circ$ corresponds to maximum shear) with the central PIL (x = 0, z = 0).

Following \cite{Leake2022}, the initial thermodynamic state is stratified using a hydrostatic atmosphere consisting of a convection zone ($z < 2.5$~Mm), photosphere/chromosphere ($2.5 < z < 10$~Mm), TR ($z=10$~Mm) and corona ($z > 10$~Mm). This thermally structured atmosphere is maintained by imposing a small background heating that decays exponentially with height,
  \begin{eqnarray}
    &Q_{\textrm{bg}}
    = Q_0 \exp(-z/ \lambda_{\textrm{bg}}).
    \label{Eqn:Qbg}
  \end{eqnarray} 
Here $Q_0=10^{-5} \textrm{Jm}^{-3}\textrm{s}^{-1}$ and a scale height of $\lambda_{\textrm{bg}}=50$~Mm are used to balance the energy losses in the TR and corona that are due to thermal conduction and optically thin radiation. 
Meanwhile, the lower part of this hydrostatic state is maintained by reducing the optically thin radiative losses to zero above the chromospheric temperature of $10^4$~K \citep{Klimchuk1987,Bradshaw2013}.

\begin{figure*}
    \includegraphics[width=\textwidth]{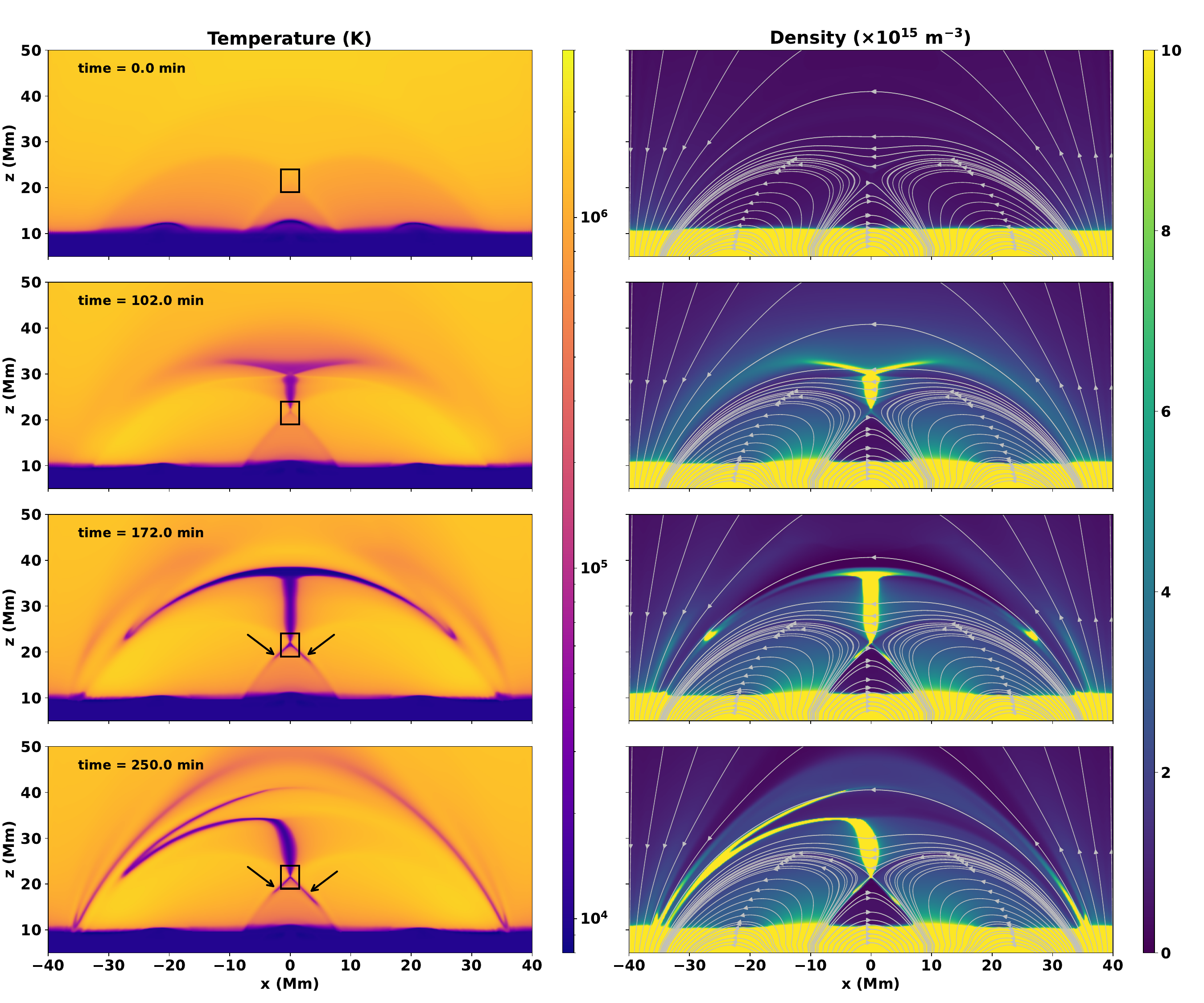}
    \caption{Time ordered snapshots of the temperature and density showing the formation of a filament and subsequent draining of hybrid filament/coronal rain condensations.
The contours are drawn according to the scales shown in the colour tables.
In the temperature plots, the black box outlines the zoomed view shown in Figure \ref{fig:nJyv_perp_x_point} and the black arrows indicate the hybrid filament/coronal rain condensations.
The grey curves on the density plots represent the magnetic field lines. 
A movie of the full time evolution of the temperature and density from $t=0-300$~min can be viewed online.
\\
    }
    \label{fig:Tn_quadrapole}
  \end{figure*}

At the left and right-hand boundaries, all velocity components and normal gradients of density, pressure, $B_y$ and $B_z$ are set to zero, while $B_x$ is determined to ensure $\nabla \cdot {\bf B}=0$. The bottom and top boundary conditions are applied at $z = 0$ and $z=80$~Mm, respectively, which are heights below and above the boundaries seen in the panels of Figure \ref{fig:Tn_quadrapole}. At the top and bottom boundaries, all velocity components and normal gradients of $B_x$ and $B_y$ are set to zero and $B_z$ is determined to ensure $\nabla \cdot {\bf B}=0$, while density and pressure are calculated to enforce hydrostatic equilibrium. 

This initial setup is out of thermodynamic equilibrium and magnetic force balance. Thus, it is first evolved for 200 minutes to a dynamic equilibrium that achieves quasi-thermal balance between the corona, TR and chromosphere, and quasi-force balance with the establishment of a steady reconnecting current sheet at the coronal X-point of the quadrapole field. Relaxation is identified as the time when only small flows, less than 5~km/s, that are associated with the steady reconnection process between the sides of the X-point remain present. In the simulation shown below, our time $t = 0$~min is taken as this end state, which has a temperature and density of approximately 1~MK and $5 \times 10^{14} \textrm{m}^{-3}$ at the coronal height where the current sheet has formed ($z=22$~Mm). 

To this relaxed system, we impose strong \lq\lq footpoint heating\rq\rq\ over a narrow height that starts at the top of the chromosphere using the same formulation as \cite{Keppens2014},
  \begin{eqnarray}
    Q_{\textrm{fp}}
    = Q_1 R(t)C(z)H(x),
    \label{Eqn:Qfp}
  \end{eqnarray} 
 where the maximum heating rate $Q_1=2 \times 10^{-3} \textrm{Jm}^{-3}\textrm{s}^{-1}$ is two orders of magnitude larger than the background value.
The temporal profile of the ramp function ($R(t)$) increases linearly from zero to one during the time interval $t=0-15$~min, and then remains steady thereafter.
Stratification of the footpoint heating takes the form,
  \begin{equation}
    C(z) = 
   \begin{cases}
      1, & \textrm{if } z \le z_{\textrm{h}}, \\
      \exp 
      \left(
       -(z - z_{\textrm{h}})^{2}/\lambda_{\textrm{h}}^{2}
      \right), & \textrm{if } z > z_{\textrm{h}},
    \end{cases}
    \label{Eqn:Qfp_z}
  \end{equation}
with $z_{\textrm{h}}=10$~Mm and $\lambda_{\textrm{h}}=5$~Mm, while the horizontal localization is given by the sum of two Gaussian peaks,
  \begin{eqnarray}
    H(x) =
    \exp 
    \left( 
    \frac{-(x + x_{\textrm{h}})^{2}}{\sigma^{2}}
    \right) 
    +
    \exp 
    \left( 
    \frac{-(x - x_{\textrm{h}})^{2}}{\sigma^{2}}
    \right),
    \hspace{0.5cm}
    \label{Eqn:Qfp_x}
  \end{eqnarray} 
with $x_{\textrm{h}}=30$~Mm and $\sigma=5$~Mm.
%
%
%
%
%
%
\section{Results} \label{sec:results}
\subsection{Magnetic \& Thermodynamic Evolution}

Figure \ref{fig:Tn_quadrapole} shows the formation and subsequent evolution of a filament above the coronal X-point in the sheared quadrapole, in response to strong localized footpoint heating. 
The two columns show contour plots of the temperature and density to illustrate the thermodynamic evolution of the plasma, while the field lines show the evolution of the magnetic field.
Each row shows a snapshot at a different time: $t = 0$~min (row 1), $t = 102$~min (row 2), $t = 172$~min (row 3) and $t = 250$~min (row 4). 
These correspond to the relaxed initial state and times during the formation of the filament and the draining of hybrid filament/coronal rain condensations. A movie of the full time evolution, using the same visualization, can be viewed online.

The heating is concentrated towards the footpoints of coronal loops \citep[e.g.,][]{Antiochos2000,Karpen2001} that connect through or above the coronal X-point. This localized energy deposition drives evaporative upflows that fill the loops with hot dense plasma, increasing the coronal density and radiative losses for the first 90~min. Eventually, the radiative losses overcome the heating source(s) at the top of the loops \citep{Antiochos1991,Antiochos1999}, and runaway cooling is triggered locally in the corona, forming a condensation \cite[]{Parker1953,Field1965} at around $t=90$~min. 
This sustained evaporation-condensation process has been termed TNE \citep{Antiochos2000,Karpen2001,Mikic2013}.
Slow and fast mode perturbations produced by this first thermal runaway then drive the growth of a thermally unstable region in the direction perpendicular to the magnetic field \citep{Fang2013,Fang2015}. The observed \lq\lq sympathetic cooling\rq\rq\ leads to coronal condensations that form quasi-simultaneously, between $t=90-130$~min, in the region above the X-point, across field lines that have different lengths.

The condensations accumulate on the marginally dipped magnetic field lines, where they are gravitationally trapped.
Collectively, these condensations form a filament above the X-point that persists from $t=130$~min.
Gravitationally unstable condensations are also continuously regenerated by TNE cycles on the arched field lines above the filament. They remain suspended in the corona only for a short period of time before draining back down to the chromosphere along the legs of the loops, resembling quiescent coronal rain \citep{Antolin2012,Auchere2018,Froment2020}. With the mass evacuated, these arched field lines then recover a hot corona, but since the strong footpoint heating remains steady evaporation soon starts again and the TNE cycle repeats \citep{Kuin1982,Froment2018,Winebarger2018,Klimchuk2019}.
Such evolution occurs around $t=140$~min, when two large coronal condensations are falling down either side of the quadrapole, while the stable filament continues to grow and starts to enhance the dips on the field lines above the X-point.

It is striking that the filament proceeds to gain enough mass in these enhanced dips to push the magnetic field to break its topology, whereupon the mass of the filament forces the field at the top and bottom of the X-point to reconnect.
This mass-driven reconnection results in the loss of filament mass through the X-point. 
These condensations drain down the field lines on the side lobes underneath the X-point, as hybrid filament/coronal rain, with multiple manifestations from $t=160$~min onwards. 
Such formation of a filament above an X-point structure followed by coronal rain falling from the magnetic dip has been reported by many observational studies of coronal cloud filaments \citep[e.g.,][]{Schad2016,Li2018,Li2019,Chen2022}. 
We note that the X-points in these observations are found at substantially higher heights ($\approx 100$~Mm) than simulated here. However, the stabilizing effect of thermal conduction is less efficient in longer loops, and so runaway cooling and the formation of condensations is even more likely to occur \citep{Klimchuk2019} in simulations of coronal cloud filaments that better match the observed heights and loop lengths.

Furthermore, even after a significant fraction of the filament has been lost, when it catastrophically drains along the magnetic field between $t=220-260$~min, mass losses through the X-point, due to mass-driven  reconnection at the X-point, continue to be observed up until the end of the simulation at $t=300$~min. These different curved drainage routes resemble the formation of \lq\lq spider legs\rq\rq\ that are sometimes associated with cloud filaments \citep{Allen1998,Schad2016}. 
We also note that the asymmetries seen in these draining dynamics are first introduced by the accumulation of numerical roundoff errors.

  \begin{sidewaysfigure*}
    \vspace*{9cm}
    \includegraphics[width=\textwidth]{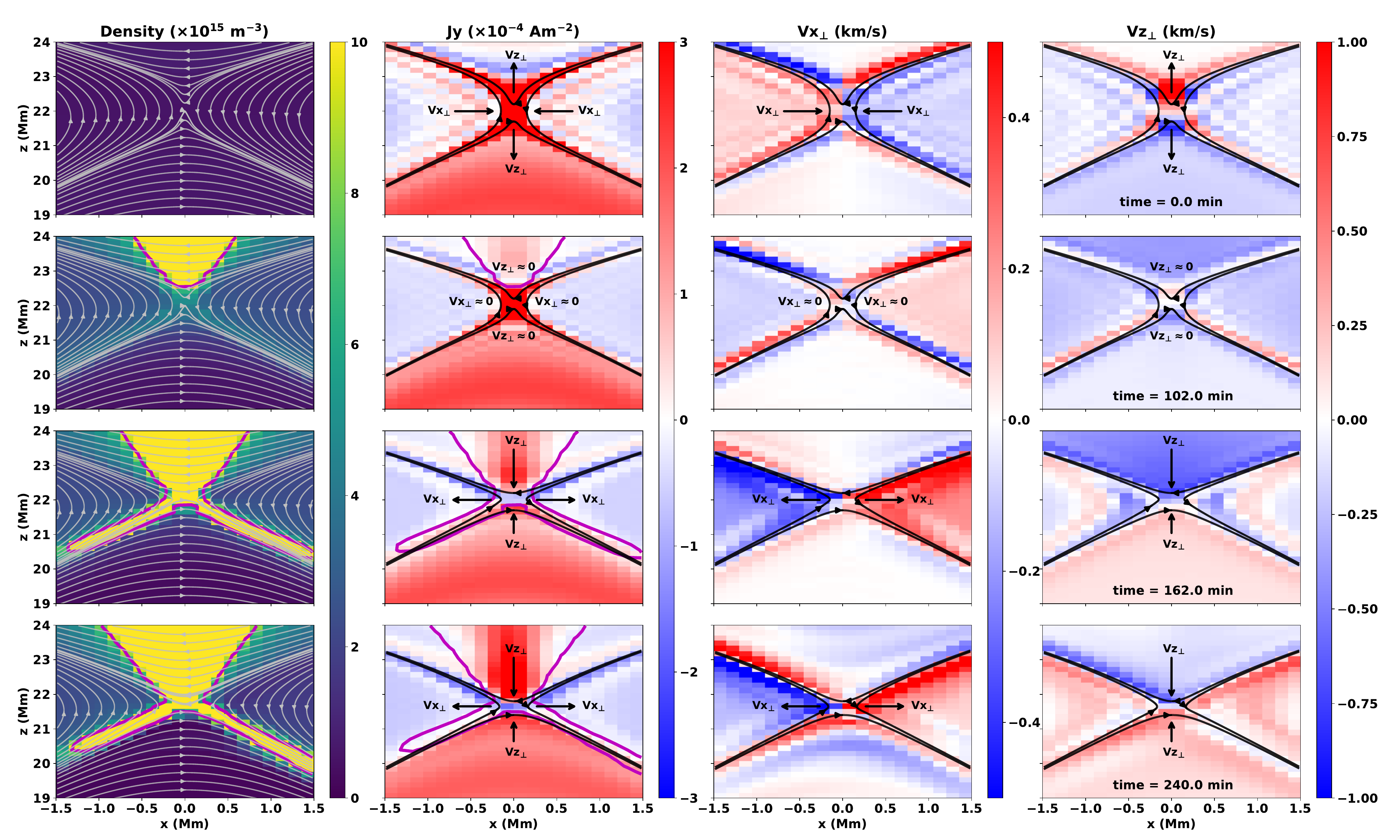}
    \caption{Zoomed view of the reconnecting current sheet at the coronal X-point during the formation of a filament and subsequent draining of hybrid filament/coronal rain condensations.
Starting from the left, the columns show time ordered snapshots of the density ($n$), $J_y$ current, and the components of the $v_x$ and $v_z$ velocities that are perpendicular to the magnetic field, denoted $v_{x_\perp}$ and $v_{z_\perp}$, respectively.
The contours are drawn according to the scales shown in the colour tables.
The grey curves on the density plots represent the magnetic field lines and the purple contour lines indicate the filament and hybrid filament/coronal rain condensations with a density value of $10^{16}$~m$^{-3}$. 
We use blacks arrows to show the direction of $v_{x_\perp}$ and $v_{z_\perp}$ at the X-point.
Red/blue in  $v_{x_\perp}$ corresponds to right/left, red/blue in $v_{z_\perp}$ corresponds to up/down, and red/blue in $J_y$ corresponds to a current directed into/out of the page.
A movie of the full time evolution of the $n$, $J_y$, $v_{x_\perp}$ and $v_{z_\perp}$ contour plots from $t=0-300$~min can be viewed online.  
    }
    \label{fig:nJyv_perp_x_point}
  \end{sidewaysfigure*}
\subsection{Signatures of Magnetic Reconnection}

This work demonstrates magnetic reconnection as a filament loss mechanism, producing hybrid filament/coronal rain via mass losses through the X-point.   
The main question that needs to be addressed about this mechanism is how does the filament force the field to reconnect at the coronal X-point? 

Figure \ref{fig:nJyv_perp_x_point} presents a detailed overview of how the mass of the filament affects the local currents at the X-point and the resulting signatures of magnetic reconnection.
The columns show a zoomed view of the density and $J_y$ current, together with $v_{x_\perp}$ and $v_{z_\perp}$, which identify reconnection inflows and outflow jets that are perpendicular to the magnetic field (separately from the field-aligned evaporation flows: 
${\bf v_\parallel = (v \cdot \hat{B}) \hat{B}}$ 
so that
${\bf v_\perp = v - v_\parallel}$).
Three main phases are used to characterize
the evolution of the reconnecting current sheet at the X-point:  Phase~1 ($t=0-70$~min), Phase~2 ($t=70-120$~min) and Phase~3 ($t=120-300$~min).
The upper two rows of Figure \ref{fig:nJyv_perp_x_point} show snapshots typical for times during the first two phases, respectively, while the lower two rows show snapshots representative of the third phase. 
These three phases can also be clearly identified throughout the bursty evolution seen in the movie online, which uses the same reconnection diagnostics. We note that this burstiness is primarily due to the intermittent nature of the reconnection at the current sheet.

Phase 1 describes the initial evolution of the reconnecting current sheet before the filament has formed.
Therefore, the filament mass plays no role in the reconnection during this first phase. 
Rather, the reconnection is characterized by a vertical current sheet that is accompanied by horizontal $v_{x_\perp}$ reconnection inflows and vertical $v_{z_\perp}$ outflow jets, which are localized at the X-point, 
as shown in the top row of Figure \ref{fig:nJyv_perp_x_point}.
Thus, the sides of the X-point are reconnecting throughout Phase 1, with $B_z$ the reconnecting field component. We also note that during this phase the reconnecting current sheet is associated with a positive $J_y$ (directed into the page) current system elongated in the vertical direction, which reinforces the magnetic structure of the quadrapole near the X-point.

The filament forms and starts to influence the reconnection at the current sheet during Phase 2. In particular, the downward gravitational force associated with the mass of the filament becomes comparable in magnitude to the upward magnetic tension force on newly reconnected field lines. 
Subsequently, as the filament gains more mass, the gravitational force begins to dominate.  
The outcome shown in the second row of Figure \ref{fig:nJyv_perp_x_point} is that downflows above the X-point related to the filament mass diminish the vertical $v_{z_\perp}$ outflow jets that are prominent throughout Phase 1. 
Therefore, the mass of the filament effectively shuts down the reconnection between the sides of the X-point during Phase 2, with a weak $J_y$ current  directed into the page forming dynamically above the X-point.

Having shut down the reconnection, the filament mass then starts to partially compress the vertical current sheet and further deform the magnetic field throughout the second half of Phase 2.
Faint asymmetric upward $v_{z_\perp}$ outflow jets seen in the movie suggest that this culminates in the generation of a series of flux ropes, each with a similar formation to that described in \cite{Keppens2014}, but here with the field pinching off multiple times and only containing a small part of the lower filament each time. 
The interpretation is that each flux rope then falls through the X-point under gravity and merges with the underlying field \citep[e.g.,][]{Liu2012,Liu2014}. 
However, the details of this process are significantly under resolved in our simulation and so we only remark that such evolution serves as a possible explanation for the filament mass on the field lines below the X-point at $t=115$~min. These details will be investigated in a future publication.

Finally, Phase 3 commences with the mass of the filament completely thinning the vertical current sheet, forcing the transition to a horizontal reconnecting current sheet with a reversal of the current at the X-point. Thereafter, the reconnection is associated with a negative $J_y$ (directed out of the page) current system elongated in the horizontal direction, which weakens the magnetic structure of the quadrapole near the X-point. This change in direction of the reconnection is confirmed by the vertical $v_{z_\perp}$ reconnection inflows and horizontal $v_{x_\perp}$ outflow jets that are shown in the lower two rows of Figure \ref{fig:nJyv_perp_x_point}. We note that downward reconnection inflows from above the X-point are generally stronger than the upward directed inflows from below. Therefore, the filament mass forces asymmetric reconnection between the top and bottom of the X-point during Phase 3, with $B_x$ the reconnecting field component.

When the field reconnects, filament mass is transferred from the dipped field lines at the top of the X-point onto the newly reconnected field lines at the sides. This transferred mass is then slowly transported outwards from the center of the reconnecting current sheet, carried on the newly reconnected field lines by the horizontal $v_{x_\perp}$ outflow jets. Due to the long transport time,  the mass initially accumulates as condensations below the X-point, where the projection of gravity along the newly reconnected field lines is small. Eventually, as they are carried further outwards, these condensations become gravitationally unstable and drain down the side lobes as hybrid filament/coronal rain. Thus, the mass-driven reconnection that characterizes Phase 3 results in the loss of filament mass from the corona to the chromosphere, via the draining of condensations through the X-point.
  \begin{figure*}
    \includegraphics[width=\textwidth]{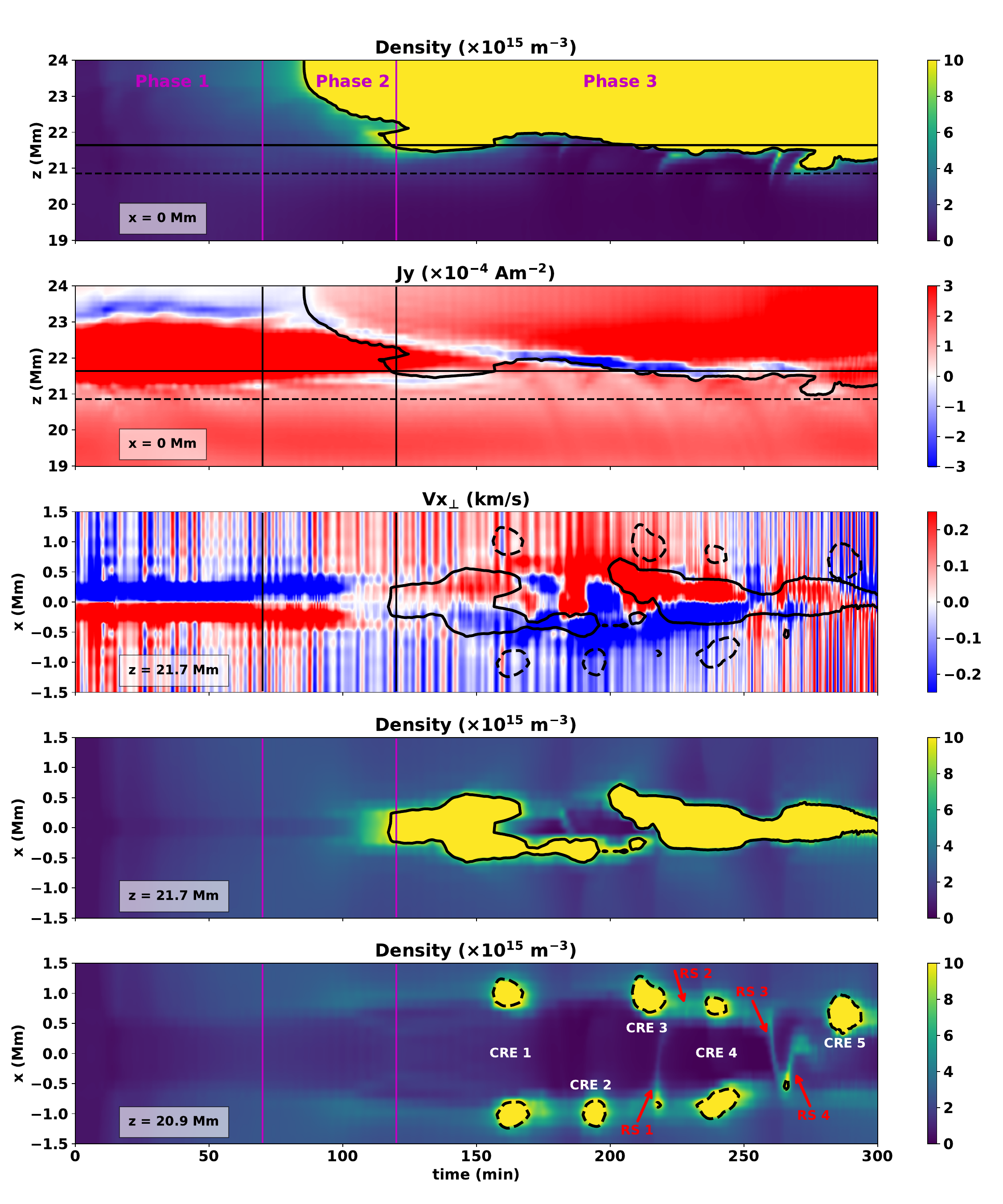}
    \caption{Signatures of reconnection, rain and rebound shocks.
\textit{Upper two rows}: Vertical dependence (vertical axis) and temporal evolution (horizontal axis) of the density and $J_y$ current at  $x = 0$~Mm.
The horizontal lines mark the two heights used in the lower three panels.
Solid vertical lines indicate the three phases of evolution associated with the reconnecting current sheet at the coronal X-point.
\textit{Lower three rows}: Horizontal dependence (vertical axis) and temporal evolution (horizontal axis) of the $v_{x_\perp}$~velocity at  $z = 21.7$~Mm, and density at $z = 21.7$~Mm and $z = 20.9$~Mm.
The solid (dashed) black contour lines indicate the filament (hybrid filament/coronal rain condensations) with a density value of $10^{16}$~m$^{-3}$.
The red arrows indicate the rebound shocks that produce clear density signatures at $z=20.9$~Mm.
\label{fig:Jynv_perp_reconnection_signatures}
}
  \end{figure*}
  \begin{figure*}
    \includegraphics[width=\textwidth]{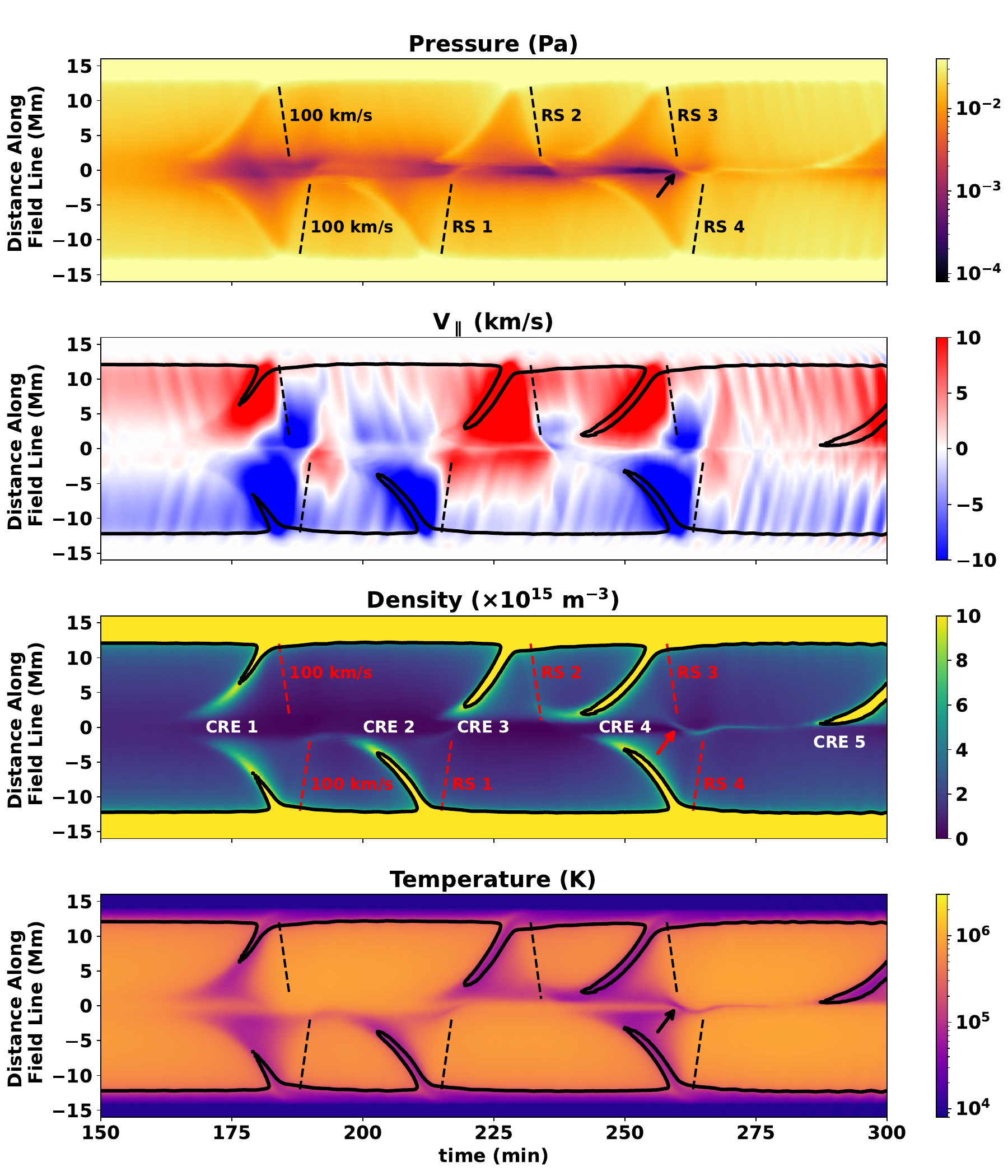}
    \caption{Time-distance plots showing the generation and subsequent evolution of rebound shocks on a field line traced through the point $(x,z) =(0,20.9) ~\textrm{Mm}$, which is located below the coronal X-point.
The plots show the dependence along the field line (vertical axis) and temporal evolution (horizontal axis) of the pressure, field-aligned velocity (${\bf v_\parallel = (v \cdot \hat{B}) \hat{B}}$), density and temperature, respectively.
The rebound shocks are the light tracks in the pressure plot that are marked by the dashed lines.
These dashed lines correspond to ${\bf v_\parallel} = 100$~km/s and are offset by around a minute so that the rebound shocks can be clearly identified. 
Consistent with Figure \ref{fig:Jynv_perp_reconnection_signatures} only the rebound shocks that produce clear density signatures at $z=20.9$~Mm are labeled (RSs 1-4).
The solid black contour lines indicate the hybrid filament/coronal rain condensations with a density value of $10^{16}$~m$^{-3}$.
The arrows indicate the condensation formed by RSs 3-4 that subsequently influence the symmetry of CRE 5.
\label{fig:rebound_signatures_v_par_pressure}}
  \end{figure*}

Some of the hybrid filament/coronal rain events observed during Phase 3 have condensations that drain down both sides of the magnetic field underneath the X-point. Such symmetric events can be seen in the lower two rows of Figure \ref{fig:nJyv_perp_x_point}. However, there are also asymmetric events, where condensations drain down on only one side of the underlying field. The explanation for these different types of rain events is discussed in the next section.

\subsection{Reconnection, Rain and Rebound Shocks}

We now turn to a detailed discussion of all the hybrid filament/coronal rain events that occur throughout the simulation. Figure \ref{fig:Jynv_perp_reconnection_signatures} presents the key reconnection diagnostics at the X-point, together with the temporal evolution of the filament and coronal rain condensations, to determine what causes the symmetric and asymmetric coronal rain events. 
The upper two rows show the time evolution of the density and $J_y$ current at the horizontal midplane ($x=0$~Mm). Rows 3 and 4 show the time evolution of the $v_{x_\perp}$ velocity and density at a coronal height that is associated with the filament ($z=21.7$~Mm). The lower row shows the evolution of the density  at a lower height that captures the draining coronal rain condensations ($z=20.9$~Mm). These rain condensations are outlined by dashed contour lines, while the filament is indicated by solid contour lines. Each plane corresponds to a particular cut from the panels shown previously in Figure \ref{fig:nJyv_perp_x_point}.

Starting with the reconnection diagnostics, the $J_y$ current and $v_{x_\perp}$ velocity clearly demonstrate the three phases that characterize the evolution of the reconnecting current sheet. In particular, the thinning and reversal of the current sheet is co-temporal with the reversal of the $v_{x_\perp}$ flows. These reversals define each of the different phases. Horizontal $v_{x_\perp}$ inflows are typical throughout Phase 1, followed by a transition period during Phase 2, before outflows dominate Phase 3. The density evolution confirms that each of these phases is directly correlated with the filament mass. Furthermore, hybrid filament/coronal rain events only occur during Phase 3, when the mass of the filament is forcing the reconnection and the resulting $v_{x_\perp}$ outflow jets can carry condensations outwards from the center of the reconnecting current sheet.

Five distinct hybrid filament/coronal rain events are identified throughout Phase 3. These coronal rain events (CREs) are labeled CRE 1-5, respectively. CREs 1 and 4 are symmetric events, with condensations draining down both sides of the magnetic field. As shown in row 3 of Figure \ref{fig:Jynv_perp_reconnection_signatures}, the symmetry for these two particular events is traced back to the filament being symmetrically located about the outflow jets in the period prior to the rain. This then allows the outflow jets to carry condensations outwards in both directions before they eventually drain down both sides.

A contrasting argument also holds for the asymmetry that is associated with CREs 2 and 3. These are asymmetric events, where a condensation drains down only on the left (CRE 2) or right-hand (CRE 3) side of the magnetic field. Prior to these rain events, the filament is found to be asymmetrically positioned within outflow jets that have a preferred direction. Thus, the condensations are directed to the left for CRE 2 and to the right for CRE 3, before proceeding to drain down that particular side. Therefore, CREs 1-4 all reveal a clear correlation between the direction of the rain condensations and the reconnection outflow jets. Specifically, a symmetric (asymmetric) positioning of the filament with respect to the outflow jets leads to symmetric (asymmetric) coronal rain events.

However, this simple interpretation is not sufficient to explain CRE 5 because this asymmetric event is traced back to a symmetric filament located about outflow jets without a preferred direction. Instead, the asymmetry in this particular event is explained by the influence of rebound shocks that are generated by compression when coronal rain condensations hit the chromosphere \citep[e.g.,][]{Fang2015,Li2022}. Recently this phenomena has been observed for the first time by \cite{Antolin2023} using the Extreme Ultraviolet Imager (EUI) on board Solar Orbiter \citep{Rochus2020}.

We highlight four different rebound shocks (RSs) in the lower row of Figure \ref{fig:Jynv_perp_reconnection_signatures}, labeled RS 1-4.
Here, the first and second rebound shocks (RSs 1-2) are produced by the impact of the condensations from CREs 2-3, while RSs 3-4 are generated by CRE 4 and are those that go on to influence the symmetry of CRE 5. Two rebound shocks are also generated by CRE 1, but these do not produce clear signatures in the density at the coronal height ($z=20.9$~Mm) considered in the lower row of Figure \ref{fig:Jynv_perp_reconnection_signatures}.

Figure \ref{fig:rebound_signatures_v_par_pressure} shows the generation and subsequent propagation of the rebound shocks on a field line below the X-point, traced through the point $(x,z) =(0,20.9) ~\textrm{Mm}$. 
The field-aligned dynamics on this traced field line are representative of the dynamics on loops just below the X-point.
This field line is traced at every snapshot to account for the compression and stretching of its length by the condensation and reconnection dynamics above.
The time-distance plots show the evolution of the pressure, field-aligned velocity (${\bf v_\parallel = (v \cdot \hat{B}) \hat{B}}$), density and temperature as functions of distance along the traced field line. 
Thus, a vertical cut in each panel represents a snapshot of the corresponding physical quantity on the traced field line.
Despite being disconnected from the filament that forms on field lines above (see e.g., Figures \ref{fig:nJyv_perp_x_point}-\ref{fig:Jynv_perp_reconnection_signatures}), the traced field line captures the draining of the hybrid filament/coronal rain condensations down the side lobes, allowing cause and effect of the rebound shocks to be identified separately from the forced reconnection dynamics.

The condensations have a high density and pressure, and fall under gravity with the draining field-aligned flows that are shown in Figure \ref{fig:rebound_signatures_v_par_pressure}.
When they drain, the condensations leave behind a low density region with low pressure at the apex of the loop. 
This creates a pressure gradient that attempts to equalize the pressure by pulling mass back via retractive upflows that start initially at the loop apex. 
When the condensations hit the chromosphere, the compression on the lower atmosphere generates a rebound shock that propagates along the field at the slow wave speed of approximately 100~km/s (corresponding to a coronal temperature of $10^6$~K).
These rebound shocks are the high pressure perturbations that are seen propagating upwards from the chromosphere in Figure \ref{fig:rebound_signatures_v_par_pressure}.
The interplay between the rebound shock waves and retractive upflows then determines the dynamics of the field-aligned flows and subsequent density evolution.

In particular, the condensation that falls down the right-hand side and generates RS 3 hits the chromosphere slightly before the condensation on the left that produces RS 4, as can be seen in the third row of Figure \ref{fig:rebound_signatures_v_par_pressure}. Consequently, the rebound shock generated on the the right (RS 3) travels upwards first and then combines with the trailing rebound shock that propagates upwards from the left (RS 4). The perturbations from these shocks followed by the increased density from the retractive upflows ($\approx 10-20$~km/s) rapidly trigger a very dynamic thermal runaway that results in the formation of a condensation.
This condensation initially forms on the right-hand side, below the X-point, before gaining further mass on the left-hand side and proceeding to propagate upwards, back to the right.
Subsequently, when it reaches the bottom of the X-point, this upward propagating condensation collides with the symmetrically positioned filament mass on newly reconnected field lines.
We note that there is no signature of this collision in Figure \ref{fig:rebound_signatures_v_par_pressure} because the traced field line does not connect through the X-point. However, the condensation forms across multiple field lines, and there is a path to collide with the filament mass on those neighboring field lines.
It is the momentum from this collision that breaks the symmetry and forces mass to drain down only on the right-hand side during CRE 5, thus demonstrating how rebound shocks and retractive upflows can also play a crucial role in determining the directionality of coronal rain events. 
%
%
%
%
\section{Discussion} \label{sec:discuss}

The major obstacle to modeling solar filaments in multi-dimensional MHD simulations is the requirement to properly resolve the TR and any dynamically forming FCTRs. 
\cite{Bradshaw2013} demonstrated that under-resolving the TR leads to coronal densities that are artificially low, which, in turn, can suppress the formation of coronal condensations that are present when the TR is properly resolved \citep{Johnston2019b}. \cite{Johnston2019b} then also went on to show that under-resolving the FCTRs can lead to discrepancies in the maintenance of condensations and their collective filament characteristics such as temperatures, densities and lifetimes. However, we have overcome these challenges, without high spatial resolution, by using the MHD TRAC method presented in \cite{Johnston2021}. This novel method makes combining the magnetic and thermodynamic
evolution of filaments in MHD models significantly more feasible moving forward.

Despite these previous difficulties, \cite{Keppens2014} were able to demonstrate that the mass of a filament can force the magnetic field to reconnect and form a flux rope. Using a similar setup, \cite{Zhou2023} went on to propose that  \lq\lq winking filaments\rq\rq\ are due to evaporation-condensation cycles, while \cite{Jervcic2024} demonstrated that different forms of localized heating lead to filaments with significantly different dynamics.
However, all three models rely on the accumulation of coronal condensations in magnetic dips above a stationary chromospheric X-point, located within the line tied lower boundary of their respective simulations. In contrast, we have simulated the formation of a filament above a dynamic coronal X-point that does not force such continuous mass accumulation in the dips. Rather condensations from the filament are allowed to drain through the coronal X-point, along reconnected field lines.
Specifically, we demonstrated three phases that characterized the evolution of the reconnecting current sheet at the X-point, culminating in filament mass losses forced by magnetic reconnection. It is striking that this reconnection was a direct consequence of the thermodynamics reversing the direction of the stabilizing current at the X-point.

Thus, the thermodynamics can drive a significant change in the magnetic structure of the corona. 
This effect cannot be captured by MHD simulations that focus only on the evolution of the magnetic structure of the filament while neglecting the thermodynamics, or by field-aligned hydrodynamic models that simulate the thermodynamic formation of the filament but neglect the evolution of the magnetic structure.

Previously, \cite{Low2012a,Low2012b} established that condensations can deform the magnetic field and fall across field lines in a resistive manner. Using 1D and 2D magnetostatic solutions, they proposed that the slippage of condensations across the field is facilitated by spontaneous formation and resistive dissipation of discrete currents. Moreover, \cite{Low2012a,Low2012b} suggested that this resistive slippage could explain the dynamic interiors of quiescent filaments \citep{Berger2008,Berger2010}. Such slippage mechanisms do not require a particular large-scale magnetic topology nor do they cause a change in the magnetic topology.

In contrast, this paper has presented a physical model that demonstrates magnetic reconnection as a filament loss mechanism, producing hybrid filament/coronal rain via mass losses through a coronal X-point. 
The formation of such a filament above an X-point structure followed by coronal rain draining from the magnetic dip has been observed in active regions \citep{Schad2016}, quiet Sun regions \citep{Li2018,Li2019} and also in regions that connect the quiet Sun to active regions \citep{Liu2012,Chen2022}. 

These observed 3D coronal configurations are typically asymmetric structures, whereas our 2.5D simulation has a highly idealized symmetry. 
\cite{Froment2018} demonstrated that the occurrence of TNE and formation of condensations in asymmetric (symmetric) loops requires asymmetric (symmetric) heating conditions. 
Furthermore, \cite{Klimchuk2019} concluded analytically that asymmetries in the heating and/or cross-sectional area must be small or offset for TNE to occur. 
However, these symmetry requirements are partially relaxed for dipped magnetic field lines because the dips can gravitationally trap the condensations. Thus, while our emphasis has been on demonstrating magnetic reconnection as a filament loss mechanism in a symmetric X-point structure, similar dynamics are also expected to be relevant for asymmetric structures that have compatible heating conditions.

In particular, our numerical simulation supports the forced reconnection scenario reported by \cite{Li2018,Li2019}. In their observations, the mass of a filament forced higher lying dipped magnetic structures to move downward and reconnect with lower lying loops, before rain condensations drained down on the legs of the newly reconnected loops. \cite{Li2018,Li2019} also proposed that the reconnection at the X-point initiated the catastrophic cooling to form the rain condensations.  However, that was not the case in our simulation. Instead, the magnetic dip acted as a reservoir of condensations that formed the filament, with TNE the dominant process invoked through evaporation-condensation that was sustained by steady footpoint heating. Reconnection was just the loss mechanism that allowed these condensations to drain through the X-point. Therefore, we conclude that while reconnection may cause magnetic dips to form and these dips can host cold condensations, it does not play an active role in triggering the thermal runaway that forms the draining condensations that fall from filaments above X-point structures. 

On the other hand, we have shown that rebound shocks produced by the impact of rain condensations on the chromosphere together with retractive upflows can cause upward propagating condensations to form on the newly reconnected loops. Consistent with recent observations \citep{Antolin2023}, the rebound shocks are magnetoacoustic waves that propagate along the field at the slow speed ($\approx 100$~km/s), causing changes to pressure and density. The retractive upflows are bounded by gravity and are thus slower ($\approx 10-20$~km/s) than the rebound shocks. However, these upflows can help extend the lifetime of the density perturbations beyond the free fall time whilst continuing to supply further mass, permitting the perturbations to grow large enough to trigger thermal runaway. This very dynamic condensation formation is associated with a fundamentally different type of thermal runaway than the thermal instability and TNE scenarios discussed by \cite{KlimchukTI2019}. Distinct from TNE, there is no strong footpoint heating that is quasi-steady on the newly reconnected loops to produce the condensation (only the small background heating term is present). In contrast with the thermal instability mechanism, the presence of the draining condensations means that the newly reconnected loops start in a very dynamic state that is violently out of thermal equilibrium and force balance.

In summary, (1) the mass of a filament can force the magnetic field to reconnect, (2) the reconnection of the magnetic field can then result in the loss of filament mass and (3) the impact of filament mass losses on the chromosphere can then cause further condensations to form along the magnetic field. All these competing effects influence the mass condensation and drainage rates of filaments. Therefore, MHD models that simulate the formation and maintenance of filaments must incorporate both the magnetic field evolution and thermodynamic response of the plasma together.
This is readily achieved now by using the MHD TRAC method developed by \cite{Johnston2020,Johnston2021}.
%
%
%
%
\section{Acknowledgments}

CDJ, PWS and MGL acknowledge support from the NASA LWS Focused Science Topic programs: NNH21ZDA001N-LWS \lq\lq The Origin of the Photospheric Magnetic Field: Mapping Currents in the Chromosphere and Corona\rq\rq (PI Pete Schuck) and NNH17ZDA001N-LWS \lq\lq Investigating Magnetic Flux Emergence with Modeling and Observations to Understand the Onset of Major Solar Eruptions\rq\rq (PI Mark Linton).
LKSD, PWS and JEL acknowledge support from the NASA GSFC Heliophysics Internal Scientist Funding Model competitive work package program (PI Rick DeVore).
MGL acknowledges support from the Office of Naval Research.
This research was supported by the International Space Science Institute (ISSI) in Bern, through ISSI International Team project \#545 (\lq\lq Observe Local Think Global: What Solar Observations Can Teach Us about Multiphase Plasmas across Physical Scales\rq\rq). This work was also supported by the Programme National PNST of CNRS/INSU co-funded by CNES and CEA.
We also thank the referee for their helpful comments that improved the manuscript.
%
%
%
\bibliography{CDJ_bibliography}{}
\bibliographystyle{aasjournal}
%
%
%
\end{document}